\documentclass[12pt]{article}
\usepackage{amsmath}
\usepackage{graphicx,psfrag,epsf}
\usepackage{booktabs}
\usepackage{multirow}
\usepackage{geometry}
\usepackage{subfigure}
\usepackage{enumerate}
\usepackage{float}
\usepackage{natbib}
\usepackage{url} 
\usepackage{color}

\newcommand{\blind}{1}

\begin{document}

\def\spacingset#1{\renewcommand{\baselinestretch}%
{#1}\small\normalsize} \spacingset{1}


\if1\blind
{
  \title{How the Avengers Assembled? \\ Analysis of Marvel Hero Social Network}
  \author{Chongyang Shi, Xuan Yu, Ziyang Ren}
  \date{}
  \maketitle
} \fi

\if0\blind
{
  \bigskip
  \bigskip
  \bigskip
  \begin{center}
    {\LARGE\bf Title}
\end{center}
  \medskip
} \fi

\bigskip
\begin{abstract}
The movies of Marvel universe are very popular among young people. Almost every young people nowadays know some of the heroes in Marvel universe, such as iron man and spider man. The data set named The Marvel Universe Social Network (MUSN) describe the social relationships of the heroes. By analyzing the MUSN, we establish a social network of Marvel heroes. We derive some basic statistics from the Marvel network, such as the number of nodes and links, the hubs, the components, the shortest path lengths and the diameter. In the next part, we analyze the structure of the Marvel network and obtain some results of the connectedness, the clustering, the degree distribution, the degree correlation. Meanwhile, we fit the power law and divide the network into different communities. In this process, we not only find that the network appears to have the small world nature, since it is obviously a scale-free network, but also find that it is very similar to the real-world social network. Further, based on the work of Loverkar et al., we do a hypothesis test on the small world nature of the network. We find that the Marvel network does have small world property. In the end, we visualize the Marvel network to give a better understanding of the network.
\end{abstract}

\noindent%
{\it Keywords: Marvel University, Scale-Free Network, Null Model, Hypotheses Testing} 
\vfill

\newpage
\spacingset{1.45} 

\section{Introduction}
The Marvel Universe is an overhead world composed of many original characters and stories based on series of comics, movies, and animations. As the Marvel series became popular around the world, more and more people became fans of these heroes, worshiping them who defend the justice of the world, and caring about their disputes or friendship.

Therefore, we collect a data set of the relationship between Marvel heroes from \textbf{kaggle} and hope to study the social network formed by them. Then compare with the real social network and explore some characteristics of the hero social network. The dataset is from Claudio Sanhueza (2016). \textit{The Marvel Universe Social Network
An artificial social network of heroes (version 1)}. Retrieved June, 2021 from \textit{https://www.kaggle.com/csanhueza/the-marvel-universe-social-network}.

\section{Basic Statistics}
\subsection{Nodes and links}

We believe that there is a connection between heroes who appear in the same comic or movie. So, there is a link between each of these heroes.

Then, in the Marvel data set, there is 6582 nodes in the Marvel Universe network and 167219 links in total. The average degree is $50.81$.

\subsection{Hubs}

By order the number of neighbors, the top 5 nodes are 
\begin{table}[h]
\begin{tabular}{|l|l|l|l|l|l|}
\hline
Nodes               & Captain America & Spiderman & Ironman & Thing & Mr.fantastic \\ \hline
Number of Neighbors & 1907            & 1737      & 1522    & 1416  & 1379         \\ \hline
\end{tabular}
\end{table}

Those 5 heroes can be regarded as the center of the 5 largest hubs. So if danger comes and you want to gather the Avengers, you can save a lot of energy by contacting these five people first.

\subsection{Component}

The Marvel network (denoted by $G_0$) is composed of 160 components in total. The largest one has $6408$ nodes and $167163$ links, while the second one has only $9$ nodes and $34$ links. The remaining 156 components are all scattered points. 

\begin{table}[h]
\centering
\begin{tabular}{|l|l|l|}
\hline
Component   & Node & link   \\ \hline
Largest one & 6408 & 167163 \\ \hline
The Secend  & 9    & 34     \\ \hline
The third   & 7    & 21     \\ \hline
The fourth  & 2    & 1      \\ \hline
5th-160th   & 1    & 0      \\ \hline
\end{tabular}
\end{table}

Thus, we find that the network consists of a very large component and many small components. This shows that a few heroes don't like to interact with other heroes. Maybe they prefer to save the world alone.

\subsection{The Shortest Path Lengths and Diameter}
Since the largest component (denoted by $G$) is not much different from the overall network $G_0$ and $G$ is connected, \textbf{from now on, we only focus on the structure and characteristics of $G$.}

The diameter of $G$ is $5$. And the average shortest path length of $G$ is $2.63843$ (as Fig.8 shows in appendix), which means each person is separated from the others by 2 people. According to the principle of Six Degrees of Separation, a certain hero can know any other hero through a maximum of three people.

\section{Structure}
\subsection{Connectedness}

Transitivity is used to measure the connectedness in this report, which measures the density of loops of length three (triangles) in a network. It is the overall probability for the network to have adjacent nodes interconnected, thus revealing the existence of tightly connected communities. As \cite{Jean} stated,

$$T=\frac{observed\;  number\;  of\;  closed\;  triplets}{ maximum\;  possible\;  number\;  of\;  closed\;  triplets\;  in\;  the\;  graph}$$

In the Marvel network, the transitivity is $0.19453$, which is quite small. There is few loop between heroes, maybe because they have to contact one-line to keep secrets.

\subsection{Clustering}

\subsubsection*{Clustering Coefficient for Iron Man}

According to the Marvel comics, Anthony Edward "Tony" Stark was a billionaire industrialist, a founding member of the Avengers, and the former CEO of Stark Industries, who have wide connections in Marvel world. Thus he is chosen for us to analyze.

Iron man has 1522 acquaints and the clustering coefficient of him is $0.062$, which reveals that there is little connection between his acquaints.

\subsubsection*{Clustering Coefficient with Degree }

\begin{figure}[h]
\centering
\includegraphics[width=0.64\textwidth]{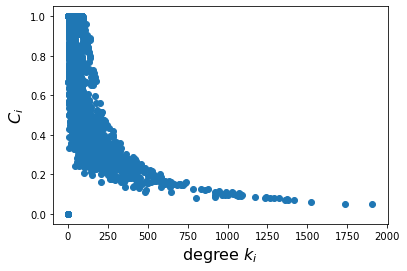}
\caption{Clustering Coefficient with Degree} 
\label{Cluster}
\end{figure}
 
  The clustering coefficient is large (close to 1) at small k, which is corresponding to the rules that the number of occurrences of hero behavior should have a certain close relationship between each stage (Fig.\ref{Cluster}).

\subsection{Degree Distribution}

\begin{figure}[h]
	\center
	\subfigure[log-log plot]{
	\includegraphics[width=0.48\linewidth]{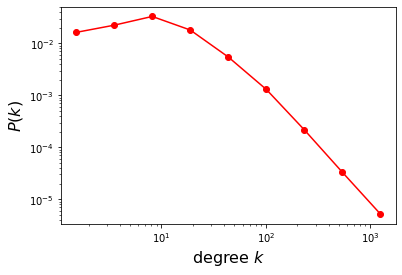}}
	\subfigure[Scatter plot]{
	\includegraphics[width=0.48\linewidth]{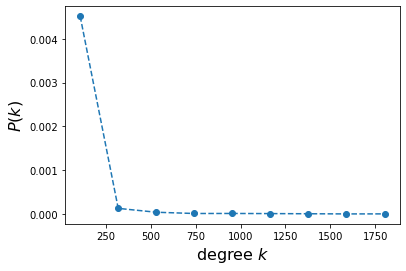}}

\subfigure[histogram]{
	\includegraphics[width=0.48\linewidth]{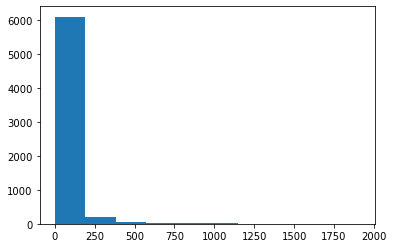}}
		
\caption{Degree Distribution}
\label{fig:1}
\end{figure}

The degree distribution of the Marvel Universe network shown in Fig.\ref{fig:1}. The observed degrees vary between k=1 (isolated nodes) and k=1908, which is the degree of the most connected node. There are also wide differences in the number of nodes with different degrees: Large amount  of the nodes have degree smaller than 100, while the $P(k)$ increase slightly at first and decline with the increase of k. This is quite similar to that of the scale free network.

\subsection{Fit the Power Law}

 In the network, we have $p_{k} \sim k^{-\gamma}$. In Marvel network, $110$ is the point from which the data displays the power-law behavior. $\gamma$ is quite close to $2.5$. Then we do a comparison between the power law and the log normal positive model, it turns out that the p-value is $1.59*10^{-13}$. The power law fits well. 
 
 The fit result can be shown in (Fig.\ref{Fit}):
 
\begin{figure}[h]
\centering
\includegraphics[width=0.6\textwidth]{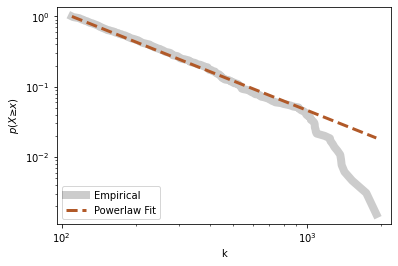}
\caption{Fit the Power Law} 
\label{Fit}
\end{figure}
 
It is obvious that there is large errors at the tail of the curve. At large k in our observation, the probability is relatively smaller than that of the prediction, which can be inferred that there are fewer heroes having wide connections than we expected.

\subsection{Community}

Greedy algorithm is an algorithm that takes the best or optimal (that is, the most advantageous) choice in the current state in each step of the selection, so as to hope that the result is the best or optimal algorithm.  In this report, the best partition found consists of the 66 communities, that is 66 groups of Marvel heroes. The modularity of this partition is $0.359$. It is a relatively clear partition.

\section{Exploration}
\subsection{Small World Property}

According to the previous analysis, we get $\gamma = 2.39$ from which we indicate that it satisfies the small world property. Thus, we do a hypothesis testing and get the average distance of the heroes is
2.638, that is,  a hero needs to contact 2 friends for a targeting strange hero.

\subsection{Null Model}

In order to test the small-world property of the network, we refer to the work of \cite{lovekar2021testing}. For a particular network, we consider the number of nodes $n$, the expected degree $2\delta$, and the mixing proportion $\beta \in [0,1]$ which is defined in the paper of \cite{lovekar2021testing}. The null model is defined as a pure Erd{\"o}s-Ren\'{y}i random graph with $n$ nodes and $p = \frac{2\delta}{n-1}$. Under this model, we define the detection of small world property as the test of the hypothesis

\[ H_0:\beta \in \{ 0,1 \} \quad v.s. \quad H_1:0<\beta<1. \]

The null hypothesis asserts that the network is either a pure ER model graph with parameters $(n, \frac{2\delta}{n - 1})$ or a pure ring lattice, while the alternative model denoted as NW-ER$(n,2\delta,\beta)$ implies the presence of significant small-world property. We define $C = \frac{3T}{3T + V}$ is the clustering coefficient or transitivity of the graph and $L$ is the average (shortest) path length of the graph. \cite{lovekar2021testing} proposed a multiple testing procedure with two test statistics $[C,L]$, which we call the intersection test. In particular we reject the null hypothesis if,

\[ \{ C > K_1 \} \cap \{ L < K_2 \} \]

for suitable choices of $K_1$ and $K_2$. The quantities $K_1$ and $K_2$ are determined by bootstrap method which involves fitting the respective null model to the observed data to estimate the parameters of the null model. Let $K_1$ be the 95th percentile of the distribution of $C$ and $K_2$ to be the 99th percentile of the distribution of $L$. We generate 100 bootstrap simulations, the result is as follows.

\begin{table}[h]
\centering
\resizebox{0.4\textwidth}{!}{%
\begin{tabular}{@{}cccc@{}}
\toprule
Model        & C        & L     & Decision                \\ \midrule
ER Network   & 0.008224 & 2.643 & \multirow{2}{*}{Reject} \\
Hero Network & 0.1945   & 2.638 &                         \\ \bottomrule
\end{tabular}%
}
\caption{The result of hypotheses testing}
\label{test}
\end{table}

From the table \ref{test}, it's clear that $C > K_1$ and $L < K_2$, we reject the null hypothesis. Thus, the Marvel network has small world property.

\subsection{Degree Correlation}
After calculate the \textbf{degree correlation coefficient}, we obtain $r=-0.1620953256887782$, in which
$$r=\frac{\sum_{jk}{jk(e_{jk}-q_jq_k)}}{\sigma^{2}_r} \qquad -1\leq r\leq1$$
And through the definition of the degree correlation coefficient,
\begin{equation*}
\begin{cases}
r\leq0 \qquad disassortative &  \\
r=0 \qquad neutral \\
r\geq0 \qquad assortative & 
\end{cases} 
\end{equation*}

\begin{figure}[h]
\centering
\includegraphics[width=0.6\textwidth]{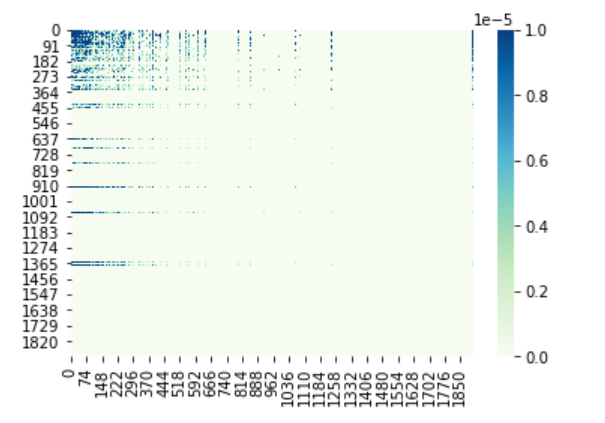}
\caption{Degree Correlation Matrix} 
\label{Degree}
\end{figure}

So the network should be disassortative (as Fig.\ref{Degree} shows). But we did not allow two nodes to have multiple edges, so we still cannot determine whether its disassortative is stuctural disassortativity caused by structural cutoff. Therefore, do \textbf{degree-preserving randomization} to decide whether the correlations observed in the Marvel network are a consequence of structural disassortativity, or are generated by some unknown process that leads to degree correlations. 

We apply degree-preserving randomization to the original network and at each step we make sure that we do not permit more than one link between a pair of nodes. \cite{Leung} stated if the real $k_{nn}(k)$ and the randomized $k_{nn}^{R-S}(k)$ are indistinguishable, then the correlations observed in a real system are all structural, fully explained by the degree distribution.

After calculation, it is found that the degree correlation coefficient of the degree-preserving randomization network is $r=-0.10581477$, which is not far from the coefficient of the original network. So, the Marvel networl is \textbf{structural disassortativity} and hubs tend to connect to small nodes.

This shows that a superhero will help those who have little influence, or a small hero will seek the help of a superhero. Of course, it may originate from that comic books about little heroes are not selling well and need to be linked with superheroes to promote sales.

\section{Visualization}
\begin{figure}[h]
\centering
\includegraphics[width=0.93\textwidth]{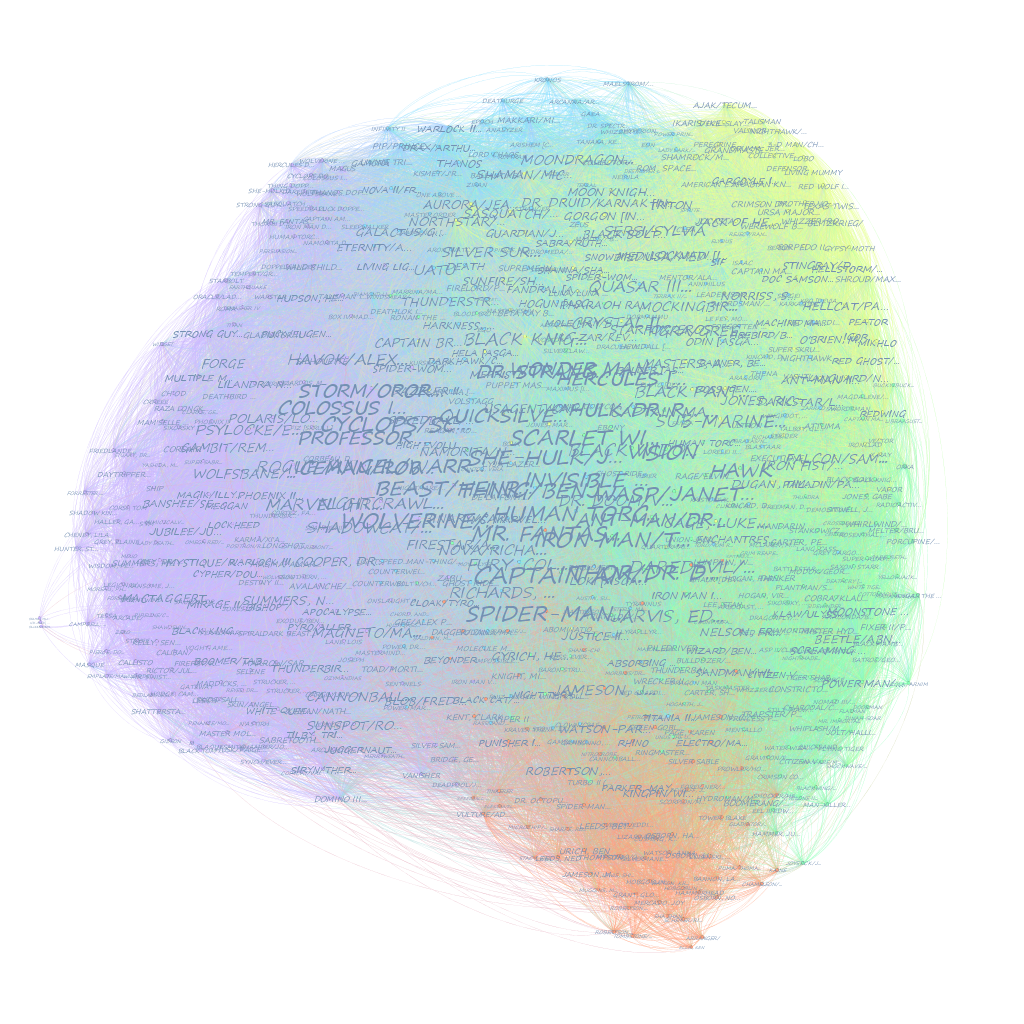}
\caption{Visualization for those nodes with $degree\geq100$ in the Marvel network} 
\label{Vis}
\end{figure}

Since the largest component is too large, only nodes with a degree greater than 100 are selected for visualization (Fig.\ref{Vis}).

For the network made by selecting nodes, it contains $715$ nodes, $54760$ edges and $5$ communities. The average degree is $153.175$, the network diameter is $3$, the average clustering coefficient is $0.601$, and the average shortest path is $1.788$. 

So in this network, each hero is separated from the others by only one hero. According to the principle of Six Degrees of Separation, a certain hero can know any other hero through a maximum of two people. This shows that many heroes are closely connected, and it is easy for them to find heroes he didn't know before through their social networks.

\section{Conclusion}

According to the previous work of \cite{Alberich} and our work on the Marvel network, we note that the Marvel network is a scale-free network which has the small world property. It is very similar to the real-world social network and the average shortest path length is only about 2.6 which means only three heroes are needed to connect with another hero for a hero. 

Recall the title \textit{How the Avengers Assembled?} Here we may get the answer to this question. Because of the small world property and the tiny average shortest path, heroes can easily contact each other. And perhaps that is why avengers could assembled in such a fast speed and defend the world’s justice.

\newpage

\providecommand{\latin}[1]{#1}
\makeatletter
\providecommand{\doi}
  {\begingroup\let\do\@makeother\dospecials
  \catcode`\{=1 \catcode`\}=2 \doi@aux}
\providecommand{\doi@aux}[1]{\endgroup\texttt{#1}}
\makeatother
\providecommand*\mcitethebibliography{\thebibliography}
\csname @ifundefined\endcsname{endmcitethebibliography}
  {\let\endmcitethebibliography\endthebibliography}{}

\bibliographystyle{achemso}

\bibliography{Reference}

\section*{Appendix}

\begin{figure}[H]
	\center
	\subfigure[Betweenness Centrality Distribution]{
	\includegraphics[width=0.44\linewidth]{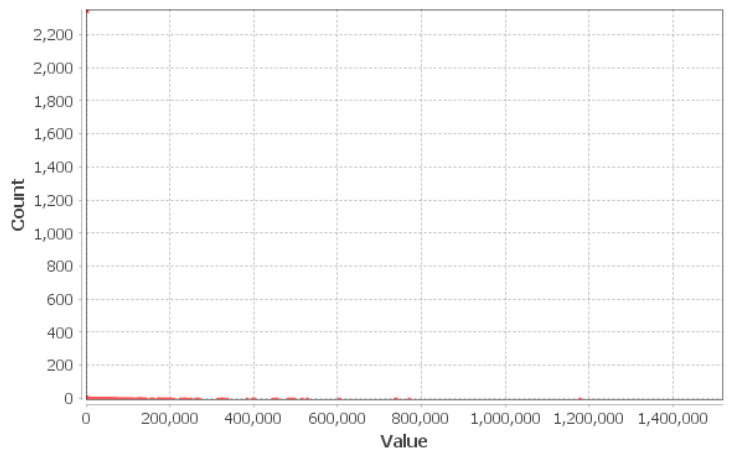}}
	\subfigure[Closeness Centrality Distribution]{
	\includegraphics[width=0.44\linewidth]{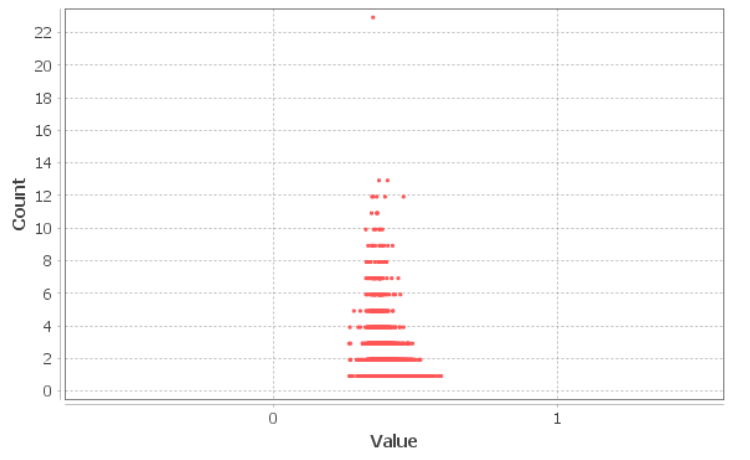}}
	
	\subfigure[Harmonic Closeness Centrality Distribution]{
	\includegraphics[width=0.44\linewidth]{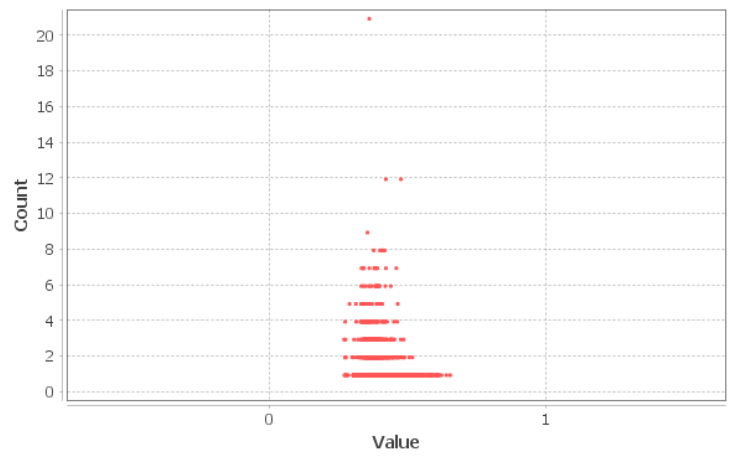}}
	\subfigure[Eccentricity Distribution]{
	\includegraphics[width=0.44\linewidth]{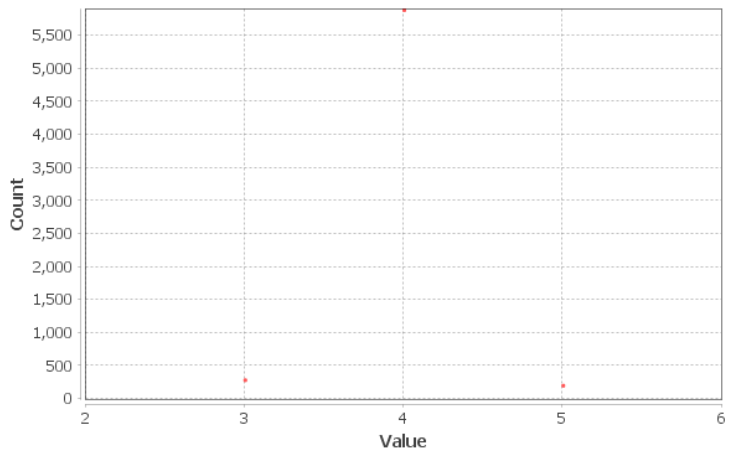}}
		
\caption{Graph Distance Report of $G$}
\label{fig:1}
\end{figure}

\end{document}